\documentclass[a4paper,fleqn,usenatbib]{mnras}
\usepackage{newtxtext,newtxmath}
\usepackage[T1]{fontenc}
\usepackage{ae,aecompl}
\usepackage{graphicx}
\usepackage{amsmath}
\usepackage{amssymb}

\newcommand\px{P_{\rm x}}
\newcommand\ax{A_{\rm x}}

\newcommand\po{P_{\rm 1O}}
\renewcommand\ao{A_{\rm 1O}}
\title[The spectral Petersen diagram]{The spectral Petersen diagram as a new tool to map pulsation modes in variable stars}
\author[Michael Hippke]{Michael Hippke$^{1}$\thanks{E-mail: michael@hippke.org}\\
$^{1}$Sonneberg Observatory, Sternwartestr. 32, 96515 Sonneberg, Germany}
\date{Accepted XXX. Received YYY; in original form ZZZ}
\pubyear{2018}
\begin{document}
\label{firstpage}
\pagerange{\pageref{firstpage}--\pageref{lastpage}}
\maketitle

\begin{abstract}
Additional pulsation modes have been discovered in many Cepheids, RR Lyrae, and other variable stars. Fourier transforms are used to find, fit and subtract the main pulsation period and its harmonics to reveal additional modes. Commonly, for every star, the strongest of these modes is presented in a ``Petersen diagram'', where the shorter-to-longer period ratio is plotted against the longer period. This diagram discards the information about temporal variations, multiple pulsation modes, and signals which are below some chosen signal to noise threshold. I here present a new tool to map pulsation modes in variable stars, dubbed the ``spectral Petersen diagram''. It shows all signals, irrespective of their multiplicity or significance. Many (thousands) of light curves can be stacked to improve the signal to noise ratio and reveal unprecedented detail about the pulsation mode space. This tool is useful to constraint the parameters of variable star models.
\end{abstract}


\section{Introduction}
Pulsating stars such as Cepheids and RR Lyrae are important variables due to their correlation between period and total light output. This correlation has allowed them to become standard candles and the first rung in the astronomical distance ladder \citep{1912HarCi.173....1L}. Before the advent of massive ground-based surveys and space-based high cadence precision photometry, they were regarded as simple, typically single-periodic, radially pulsating stars.

\begin{figure}
\includegraphics[width=\linewidth]{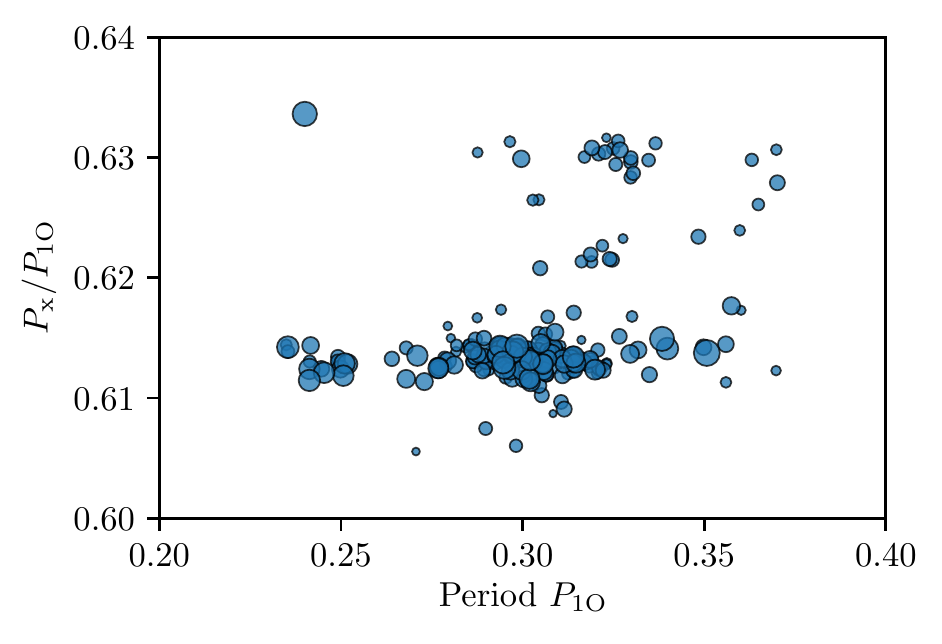}
\caption{\label{fig:classical_petersen}Petersen diagram of 260 RRc stars for $\px/\po\in(0.6, 0.64)$ and $\po\in(0.2, 0.4)\,$d. In contrast to the classical plot, here the symbol sizes represent the amplitude ratios $\ax / \ao$. Data collated from \citet{
2015MNRAS.451L..25N,
2015MNRAS.453.2022N,
2009A&A...494L..17O,
2009AcA....59....1S,
2012MNRAS.424.2528S,
2014A&A...570A.100S,
2015MNRAS.447.2348M,
2007MNRAS.379.1498G,
2012A&A...540A..68C,
2015ApJS..219...25J}.}
\end{figure}

The ``Petersen Diagram Revolution'' \citep{2017EPJWC.15206003S} resulted in a large extension of the known, and in the discovery of new forms of multiperiodic pulsation. Among Cepheids, the most important additional variability is found in the $\px/ \po\in(0.6, 0.64)$ range, assumed to be caused by non-radial pulsation. Similar signals have been found among RR Lyrae stars. The diagram by \citet{1973A&A....27...89P} is a widely used scatterplot to show the shorter-to-longer period ratio versus the longer period (Figure~\ref{fig:classical_petersen}). It has the disadvantage, however, that only one
\citep[e.g.,][]{2018MNRAS.478.1425S,2018A&A...610A..86S} or a few \citep[e.g.,][]{2015MNRAS.453.2022N} shorter periods are presented per star. Over the last years, however, it has become clear that many stars exhibit several weaker modes. In some cases, these pulsations drift and/or switch over time. Putting only one signal in a diagram discards a lot of information. Furthermore, one has to choose some signal to noise threshold above which a signal is considered a detection.

In this paper, I present a novel method, dubbed the ``spectral Petersen diagram''. It allows to show multiple secondary periods, irrespective of their significance. Through stacking of many such spectra, weak underlying structure in the Petersen diagram may be revealed. Section~\ref{sec:sft} explains the method, followed by an application to RR Lyrae stars in section~\ref{sec:application} and a discussion of its (dis)advantages in section~\ref{sec:discussion}.

%\pagebreak

\begin{figure*}
\includegraphics[width=\linewidth]{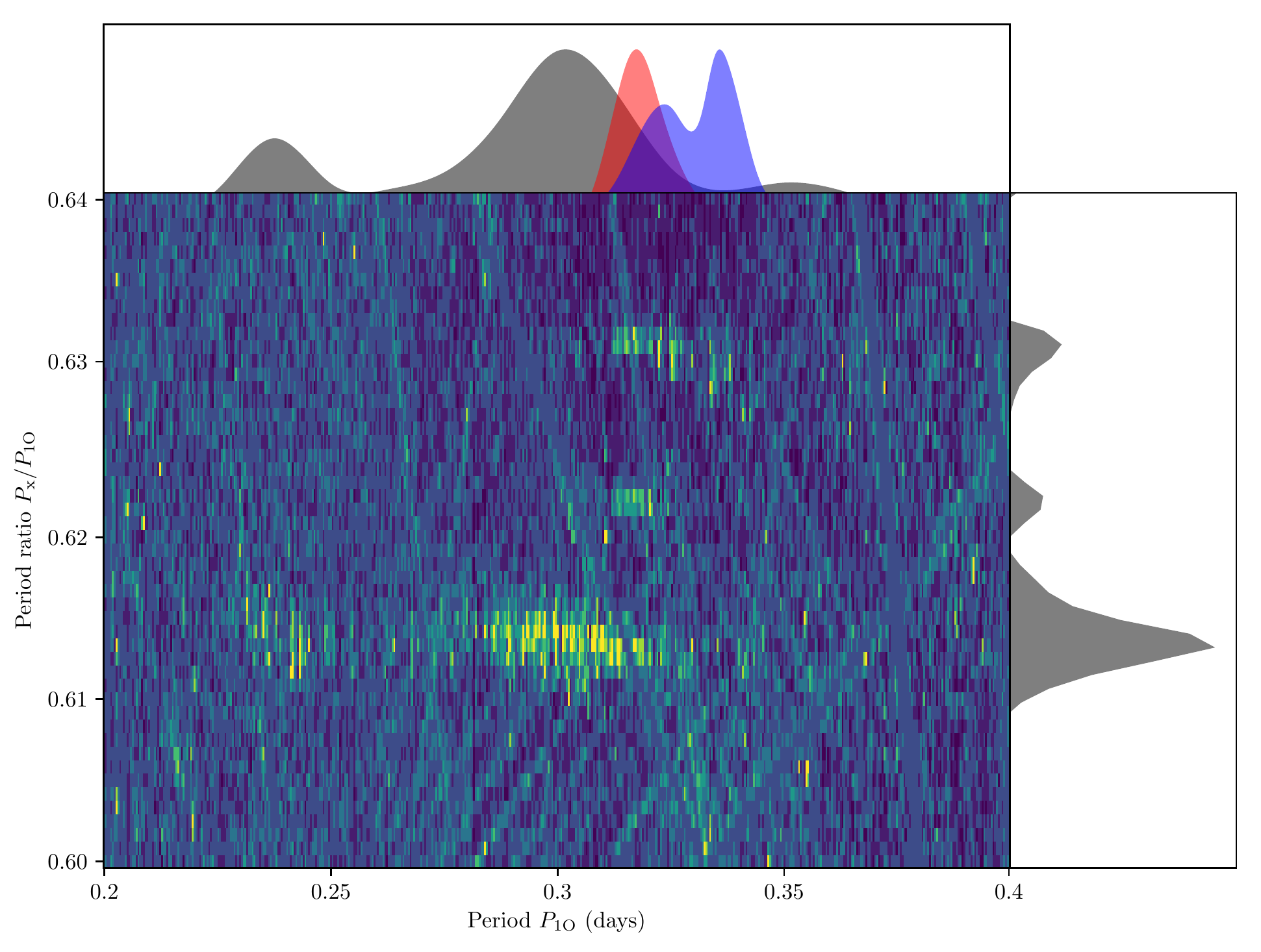}
\caption{\label{fig:spectral_map}Spectral Petersen diagram for RRc stars in $\px/\po\in(0.6, 0.64)$ and $\po\in(0.2, 0.4)\,$d. Brighter colors represent higher signal levels. Top: Probability distributions for the three signal segments, smoothed with splines of $\Delta P_{\rm 1O}=0.01\,$d. The signals near 0.613, 0.622, 0.63 are shown as gray, red, and blue colors. Right: Probability distribution over the summed $P_{\rm x}/P_{\rm 1O}$ axis.}
\end{figure*}

\begin{figure}
\includegraphics[width=\linewidth]{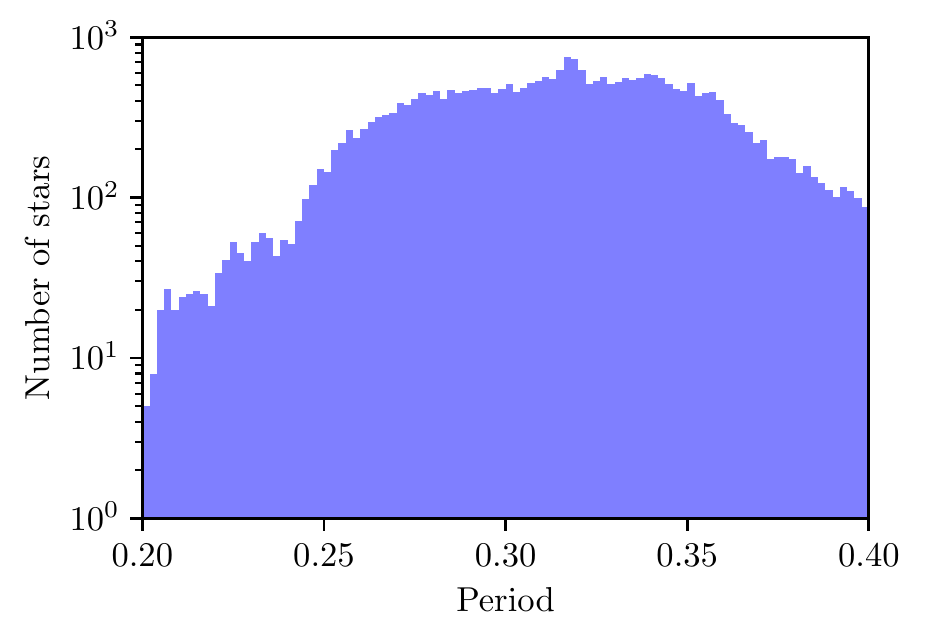}
\caption{\label{fig:histo}Histogram of the number of stars used in Figure~\ref{fig:spectral_map} as a function of $P_{\rm 1O}$.}
\end{figure}

\section{The method of Spectral Fourier Transforms}
\label{sec:sft}
Averaged Fourier spectra of RRc stars have already been shown in \citet{2003ApJ...598..597A,2018MNRAS.tmp.1796N}, but in one-dimensional form summed over all periods. As follows I extend this method to a two-dimensional graph covering the $P_1$ and $P_{\rm x}/P_{1}$ space.

Fourier spectra are generated for each star separately, sampled at the Nyquist rate. The main pulsation frequency and its harmonics are fitted and subtracted out, following the usual procedures \citep[see e.g.,][]{2003A&A...398..213M,2007MNRAS.379.1498G} with tools such as \texttt{Period04} \citep{2004IAUS..224..786L}, \texttt{SigSpec} \citep{2007A&A...467.1353R}, \texttt{CLEANest} \citep{1995AJ....109.1889F} or \texttt{TiFrAn} \citep{2004ESASP.559..396C}. After this prewhitening of $P_1$ and its harmonics, it is common to find that significant residuals remain, caused by long-term period variations. These could be reduced with time-dependent prewhitening \citep{2015MNRAS.447.2348M,2018A&A...610A..86S,2018MNRAS.478.1425S}, but require a choice for the light curve segment length (e.g., one season). Also, daily aliases may remain with power comparable to secondary pulsations. If not removed completely, residuals can be masked during further processing.

All cleaned spectra are then linearly resampled to a common resolution (e.g., $R = 10{,}000$) in the $\px/ P_1$ region of interest. Then, they are co-added in bins of the main pulsation period. A larger number of bins increases the resolution (in $P_1$) at the expense of noise, and vice versa. Weights per stellar spectrum are given based on the square root of the number of photometric data points. Finally, each bin is normalized based on its median to remove the issue that bins with a higher number of observations (and/or more data points) appear brighter.

The resulting matrix can now be plotted so that the ordinate and the abscissa follow the convention of the usual ``Petersen diagram'', but each point in the matrix is represented as a pixel with a brightness based on its weighted, stacked, and normalized sum. It is useful to down-sample the spectral resolution (e.g., to $R=1{,}000$) for printing purposes. This is the spectral Petersen diagram.

\begin{figure}
\includegraphics[width=\linewidth]{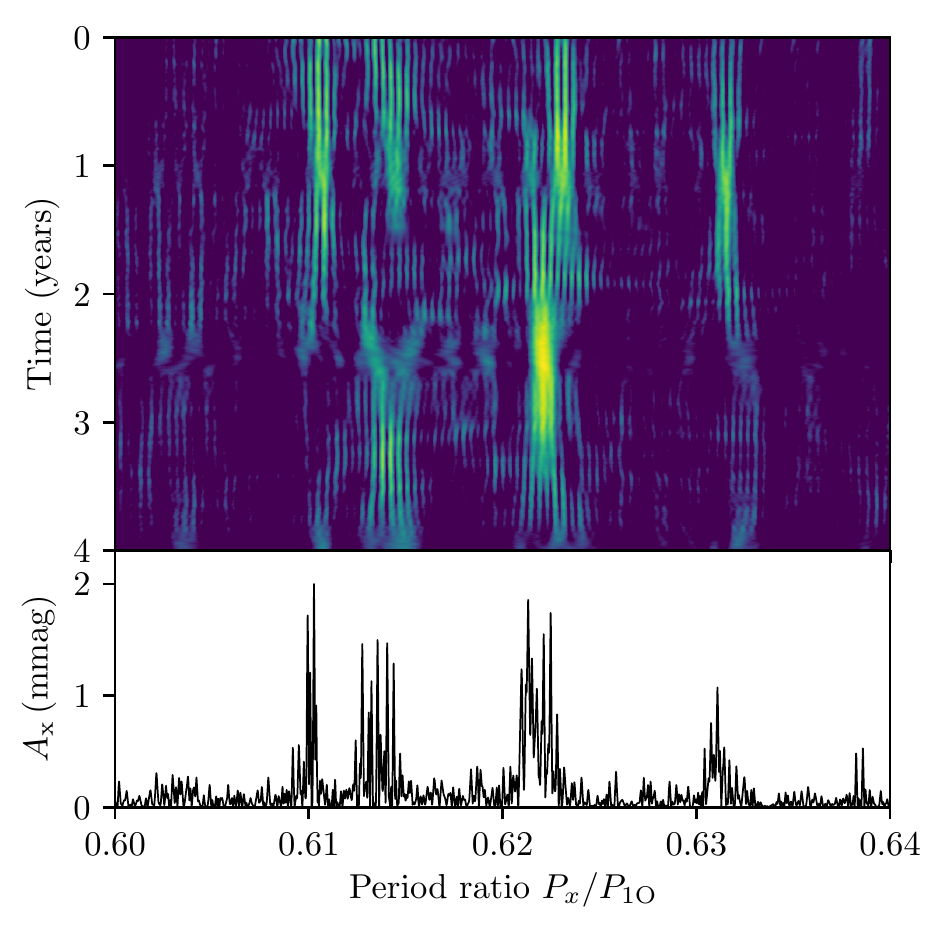}
\caption{\label{fig:time_map}Top: Temporal evolution of OGLE-BLG-RRLYR-07806 over four observation seasons. The periodograms are built using a sliding boxcar with a 50\,d window. The duration of the signals is of order one year. Artifacts are due to the interpolation over seasonal data gaps. Bottom: Prewhitened Fourier spectrum over the total data set of this star. From this classical spectrum, only the highest peak (near $P_{\rm x}/P_{\rm 1O}=0.61$) would appear in the Petersen diagram, while the other signals are discarded.}
\end{figure}

\section{Application to the 0.6-modes in RRc stars}
\label{sec:application}
As an exemplary application of the spectral Petersen diagram, I use RRc light curves from the Optical Gravitational
Lensing Experiment (OGLE) phases III \citep{2008AcA....58...69U} and IV \citep{2015AcA....65....1U} as well as variables identified in the Catalina Sky Survey \citep[SSS,][]{2009ApJ...696..870D} by \citet{2014ApJS..213....9D,2017MNRAS.469.3688D}.

Data from more surveys exist, but were not found to be useful. The Quest RR Lyrae survey identified 365 RRc stars in the northern  \citep{2004AJ....127.1158V,2006AJ....132..714V,2012MNRAS.427.3374M} and 359 in the southern hemisphere \citep{2014ApJ...781...22Z}, but the small number of observations (typically 20) per star does not allow to search for secondary pulsation modes. The VVV Survey focused on RRab stars (and identified 960) as these are easier to classify, but excluded RRc stars from their search \citep{2017AJ....153..179M,2018arXiv180704303C}. The same is true for the SuperWASP catalogue of RR Lyr stars \citep{2017A&A...607A..11G}, and the MACHO survey \citep{2000ApJ...542..257A,2017A&A...607A..11G}. The PanSTARRS classification work is still in progress \citep{2018arXiv180310028J}. Gaia DR2 identified $228{,}904$ RR Lyrae from the all-sky classifier and the SOS pipeline combined, but has typically only 15--100 observations per star \citep{2018arXiv180502079C,2018arXiv180409373H}. Similar issues exist for the LINEAR \citep{2013AJ....146...21S} and EROS \citep{2014A&A...566A..43K} surveys.

From OGLE and SSS, a total of $\sim1.9\times10^7\,$ photometric data points were processed, covering $35{,}096$ light curves of RRc stars in the galactic bulge, LMC, and SMC with up to 3 light curves of the same object. Individual light curves of the same objects were treated separately to avoid differences in zeropoints and filters. Observations span more than a decade.
Fourier spectra were generated for a region of particular interest in $\px/\po\in(0.6, 0.64)$ and $\po\in(0.2, 0.4)\,$d. RRc stars have been found to often exhibit multiple peaks in this region \citep[e.g.,][]{2015MNRAS.447.1173N,2016MNRAS.461.2934S,2017MNRAS.467.2349S}. These period ratios have been speculated to arise due to strange nonchaotic dynamics \citep{2015ApJ...798...42H,2015PhRvL.114e4101L,2016PhyD..316...16L}

%Spectra were cross-checked to be virtually identical to the published values in \citet{2015MNRAS.453.2022N}.
After stacking, the spectral Petersen diagram (Figure~\ref{fig:spectral_map}) reveals three distinct regions of power excess, coincident with the individual detections shown in Figure~\ref{fig:classical_petersen}.

The average spectral power of the signals is not immediately apparent due to the normalization. It can be estimated by co-adding without normalization and dividing by the applied weights. The noise floor in Figure~\ref{fig:spectral_map} corresponds to brightness variations of $\sim 0.1\,$mmag, and the strongest peaks are $\sim 1\,$mmag. The color stretch is linear. Clearly, these secondary pulsations are minuscule.

Figure~\ref{fig:histo} shows a histogram of the number of stars used in the spectral Petersen diagram. The highest numbers of stars per bin are found in the range $0.27 \lesssim P_{\rm 1O} \lesssim 0.36\,$d, covering the strongest secondary pulsations. Yet, at $P_{\rm 1O}\sim0.24\,$d, some signal power is seen, where the number of stars per bin is lower by an order of magnitude.

Future work can map out a larger area of period ratios for different sub-types of variable stars (e.g., Cepheids, $\delta$ Sct, RRd stars), and differentiate between metallicities, locations, or other factors.

\section{Discussion of strengths and weaknesses}
\label{sec:discussion}

\subsection{Visibility of weak signals and sensitivity to temporal evolution}
The classical ``Petersen diagram'' shows at most a few peaks per star. It is the right tool for cases where constant, single, individually significant detections are present. As shown by \citet{2015MNRAS.453.2022N}, however, stars often exhibit multiple peaks in the region, and their strength changes over the course of months or years. The attribution of one star to its highest peak, e.g. at 0.615, is therefore a snapshot in time, and discards all other peaks. I show this time evolution using a sliding peridograms for OGLE-BLG-RRLYR-07806 (Figure~\ref{fig:time_map}).

The spectral Petersen diagram includes all peaks, irrespective of their statistical significance. Consequently, a ``0.61-star'' can also contribute power at 0.62 etc., and vice versa. It is the right tool for cases where many stars exhibit (single or multiple) weak signals, which are at or below some sensible significance limit. Many spectra can be stacked to reduce the averaged noise and reveal otherwise invisibly weak signals. The method of stacking noisy data has been very successful in the detection of transits from primary and secondary exoplanet eclipses \citep{2014ApJ...794..133S,2017AJ....154..160S}, exomoons \citep{2015ApJ...806...51H,2018AJ....155...36T}, or exotrojans \citep{2015ApJ...811....1H}. It may prove equally useful for variable stars.

\citet{2015MNRAS.453.2022N} found 260 stars with significant ($4\,\sigma$) signals in $\px/\po\in(0.6, 0.64)$. When removing these stars from the data, the spectral map still shows the 0.61 peaks at low signal to noise ratio, with powers of $\sim0.2\,$mmag. The weaker 0.62 and 0.63 islands disappear in the noise floor, and are thus $\lesssim0.2\,$mmag on average. It remains unknown whether there is a large number of stars (e.g., 50\,\%) which exhibit no secondary pulsations, or whether it is present in all stars at very low (sub mmag) levels. Using space-based Kepler photometry, apparently non-modulated (at sub mmag amplitude limits) RRc stars have been found \citep{2011MNRAS.417.1022N,2013ApJ...773..181N,2018MNRAS.473..412B}.

The method of spectral Fourier transforms could also be used in the future to study RRab stars. Based on Kepler photometry, it has been argued that essentially all RRab-stars (of 151 tested) are Blazhko-modulated \citep{1907AN....175..325B,2018A&A...614L...4K}. In 40 per cent of the stars, the amplitude is $<2\,$mmag, thus buried in noise for most ground-based individual light curves \citep{2009MNRAS.400.1006J}. It is, however, very difficult to disentangle instrumental noise present in Kepler light curves. Such artifacts may arise from the assigned aperture unique to the star, and can thus not completely be removed using cotrending vectors \citep{2017EPJWC.16004009P}. The analysis of RRab stars using the spectral Fourier transforms is promising as a fraction of RRab stars also show a series of different additional modes and peaks which seem to be connected to the presence of the Blazhko effect \citep{2010MNRAS.409.1585B,2014ApJS..213...31B,2014A&A...570A.100S,2017EPJWC.16004008M}.

\subsection{Computational effort}
The creation of a stacked map requires additional computational effort. The underlying creation of Fourier transforms is essentially identical to the classical analysis. Calculating Fourier transforms, including fitting and subtracting the main pulsation period and its harmonics, takes a few seconds per light curve, using one 4\,GHz core. Fourier transforms for a dataset like OGLE-IV ($\approx10^4$ spectra with $\approx10^7$ data points) can thus be created in less than one hour. Resampling, co-adding, and normalizing these spectra takes $\sim1\,$hr. Thus, the computational effort is increased by a factor of two, but remains small.

\subsection{False-positive detections from aliases}
The blanked lines in Figure~\ref{fig:spectral_map} arise from aliases and period variations. The prewhitening removed these only imperfectly, as no individual time-dependent prewhitening was used. As can be seen, some lines intersect with the regions of interest, i.e. daily aliases might be mistaken for secondary pulsations. We checked the literature for such cases and found several stars that might in fact be false positives. One such example is OGLE-LMC-11983, identified by  \citet{2009AcA....59....1S} as a 0.61-star, which lies exactly at an alias. While such identification can be done using pure Fourier transforms, the 2D map makes false positives candidates immediately obvious.

\subsection{Data required to make a useful spectral Petersen diagram}
The sensitivity to weak signals is a function of many factors. At first approximation, the noise floor $N$ reduces with the square root of the number of stacked spectra $S$. This assumes that all spectra are of the same quality.

Similarly, the noise per spectrum reduces with the square number of data points $P$. This assumes that all data points have about the same photometric uncertainty. Combining these two factors, the noise floor scales as $N \propto \sqrt{S \sqrt{P}}$. Other contributing factors include the photometric errors and the sampling frequency. For example, the periods of many RR Lyrae are near integer multiples of daily aliases (e.g., $1/3\,$d), resulting in systematic noise hard to remove completely.

Adding more data from the surveys described in section~\ref{sec:application}, and future surveys \citep{2018arXiv180310026M} such as
LSST \citep{2002SPIE.4836...10T},
TESS \citep{2015JATIS...1a4003R}, and
PLATO \citep{2014ExA....38..249R},
will increase the signal to noise ratio by a factor of a few.

\section{Conclusion and outlook}
\label{sec:conclusion}
``If all you have is a hammer, everything looks like a nail'' \citep{kaplan1964conduct}. Extending the astronomer's toolbox is surely beneficial, although the usefulness of any new tools has to be shown in practice over time. The spectral Petersen diagram can reveal weak, individually insignificant pulsation modes through stacking of multiple stars. It could also help to enlighten us about the underlying physical origin of the these modes. Knowing more about their shape and prevalence can guide theorists to constraint and adjust their models more precisely to reflect reality. For example, the minimal slope of the 0.61 signals (only slightly higher $P_{\rm x}/P_{\rm 1O}$ for shorter $P_{\rm 1O}$) requires a low average metallicity \citep[compare Figure 4 in][]{2016CoKon.105...23D}. On the other hand, the 0.63 signals appear to drop in $P_{\rm x}/P_{\rm 1O}$ for $P_{\rm 1O}>0.33\,$d, which is not predicted by the \citet{2016CoKon.105...23D} model. New data will be required to show whether this effect is real. Probing the correlation between metallicity in RRc and their exact secondary periodicity could verify or falsify pulsation models. Some questions might remain: Why is the lower sequence in RRc stars so much denser populated? This is not seen in Cepheids, which exhibit similar secondary modes \citep{2008CoAst.157..343M,2009MNRAS.394.1649M,2016MNRAS.458.3561S}.

The spectral Petersen diagram can also be used to examine other regions of interest, such as RR Lyrae with period ratios near 0.68-0.72 \citep{2015MNRAS.451L..25N,2017MNRAS.465.4074P}, Cepheids \citep{2018MNRAS.478.1425S}, and even RRab stars \citep{2016MNRAS.461.2934S,2017EPJWC.16004008M}. High quality Kepler K2 photometry can be employed to shed new light on very weak pulsations \citep{2015MNRAS.452.4283M,2017EPJWC.16004008M,2017EPJWC.16004009P,2018arXiv180705369P}.

\section*{Acknowledgements}
The CSS survey is funded by the National Aeronautics and Space Administration under Grant No. NNG05GF22G issued through the Science Mission Directorate Near-Earth Objects Observations Program.

\bibliographystyle{mnras}
\bibliography{references}

\begin{thebibliography}{}
\makeatletter
\relax
\def\mn@urlcharsother{\let\do\@makeother \do\$\do\&\do\#\do\^\do\_\do\%\do\~}
\def\mn@doi{\begingroup\mn@urlcharsother \@ifnextchar [ {\mn@doi@}
  {\mn@doi@[]}}
\def\mn@doi@[#1]#2{\def\@tempa{#1}\ifx\@tempa\@empty \href
  {http://dx.doi.org/#2} {doi:#2}\else \href {http://dx.doi.org/#2} {#1}\fi
  \endgroup}
\def\mn@eprint#1#2{\mn@eprint@#1:#2::\@nil}
\def\mn@eprint@arXiv#1{\href {http://arxiv.org/abs/#1} {{\tt arXiv:#1}}}
\def\mn@eprint@dblp#1{\href {http://dblp.uni-trier.de/rec/bibtex/#1.xml}
  {dblp:#1}}
\def\mn@eprint@#1:#2:#3:#4\@nil{\def\@tempa {#1}\def\@tempb {#2}\def\@tempc
  {#3}\ifx \@tempc \@empty \let \@tempc \@tempb \let \@tempb \@tempa \fi \ifx
  \@tempb \@empty \def\@tempb {arXiv}\fi \@ifundefined
  {mn@eprint@\@tempb}{\@tempb:\@tempc}{\expandafter \expandafter \csname
  mn@eprint@\@tempb\endcsname \expandafter{\@tempc}}}

\bibitem[\protect\citeauthoryear{{Alcock} et~al.,}{{Alcock}
  et~al.}{2000}]{2000ApJ...542..257A}
{Alcock} C.,  et~al., 2000, \mn@doi [\apj] {10.1086/309530}, \href
  {https://ui.adsabs.harvard.edu/#abs/2000ApJ...542..257A} {542, 257}

\bibitem[\protect\citeauthoryear{{Alcock} et~al.,}{{Alcock}
  et~al.}{2003}]{2003ApJ...598..597A}
{Alcock} C.,  et~al., 2003, \mn@doi [\apj] {10.1086/378689}, \href
  {https://ui.adsabs.harvard.edu/#abs/2003ApJ...598..597A} {598, 597}

\bibitem[\protect\citeauthoryear{{Benk\H{o}}}{{Benk\H{o}}}{2018}]{2018MNRAS.473..412B}
{Benk\H{o}} J.~M.,  2018, \mn@doi [\mnras] {10.1093/mnras/stx2338}, \href
  {https://ui.adsabs.harvard.edu/#abs/2018MNRAS.473..412B} {473, 412}

\bibitem[\protect\citeauthoryear{{Benk\H{o}} et~al.,}{{Benk\H{o}}
  et~al.}{2010}]{2010MNRAS.409.1585B}
{Benk\H{o}} J.~M.,  et~al., 2010, \mn@doi [\mnras]
  {10.1111/j.1365-2966.2010.17401.x}, \href
  {https://ui.adsabs.harvard.edu/#abs/2010MNRAS.409.1585B} {409, 1585}

\bibitem[\protect\citeauthoryear{{Benk\H{o}}, {Plachy}, {Szab{\'o}},
  {Moln{\'a}r}  \& {Koll{\'a}th}}{{Benk\H{o}}
  et~al.}{2014}]{2014ApJS..213...31B}
{Benk\H{o}} J.~M.,  {Plachy} E.,  {Szab{\'o}} R.,  {Moln{\'a}r} L.,
  {Koll{\'a}th} Z.,  2014, \mn@doi [The Astrophysical Journal Supplement
  Series] {10.1088/0067-0049/213/2/31}, \href
  {https://ui.adsabs.harvard.edu/#abs/2014ApJS..213...31B} {213, 31}

\bibitem[\protect\citeauthoryear{{Bla{\v{z}}ko}}{{Bla{\v{z}}ko}}{1907}]{1907AN....175..325B}
{Bla{\v{z}}ko} S.,  1907, \mn@doi [Astronomische Nachrichten]
  {10.1002/asna.19071752002}, \href
  {https://ui.adsabs.harvard.edu/#abs/1907AN....175..325B} {175, 325}

\bibitem[\protect\citeauthoryear{{Chadid}}{{Chadid}}{2012}]{2012A&A...540A..68C}
{Chadid} M.,  2012, \mn@doi [\aap] {10.1051/0004-6361/201117408}, \href
  {https://ui.adsabs.harvard.edu/#abs/2012A&A...540A..68C} {540, A68}

\bibitem[\protect\citeauthoryear{{Clementini} et~al.,}{{Clementini}
  et~al.}{2018}]{2018arXiv180502079C}
{Clementini} G.,  et~al., 2018, preprint, \href
  {http://adsabs.harvard.edu/abs/2018arXiv180502079C} {} (\mn@eprint {arXiv}
  {1805.02079})

\bibitem[\protect\citeauthoryear{{Contreras Ramos} et~al.,}{{Contreras Ramos}
  et~al.}{2018}]{2018arXiv180704303C}
{Contreras Ramos} R.,  et~al., 2018, preprint, \href
  {http://adsabs.harvard.edu/abs/2018arXiv180704303C} {} (\mn@eprint {arXiv}
  {1807.04303})

\bibitem[\protect\citeauthoryear{{Csubry} \& {Koll{\'a}th}}{{Csubry} \&
  {Koll{\'a}th}}{2004}]{2004ESASP.559..396C}
{Csubry} Z.,  {Koll{\'a}th} Z.,  2004, in SOHO 14 Helio- and Asteroseismology:
  Towards a Golden Future. p.~396

\bibitem[\protect\citeauthoryear{{Drake} et~al.,}{{Drake}
  et~al.}{2009}]{2009ApJ...696..870D}
{Drake} A.~J.,  et~al., 2009, \mn@doi [\apj] {10.1088/0004-637X/696/1/870},
  \href {https://ui.adsabs.harvard.edu/#abs/2009ApJ...696..870D} {696, 870}

\bibitem[\protect\citeauthoryear{{Drake} et~al.,}{{Drake}
  et~al.}{2014}]{2014ApJS..213....9D}
{Drake} A.~J.,  et~al., 2014, \mn@doi [The Astrophysical Journal Supplement
  Series] {10.1088/0067-0049/213/1/9}, \href
  {https://ui.adsabs.harvard.edu/#abs/2014ApJS..213....9D} {213, 9}

\bibitem[\protect\citeauthoryear{{Drake} et~al.,}{{Drake}
  et~al.}{2017}]{2017MNRAS.469.3688D}
{Drake} A.~J.,  et~al., 2017, \mn@doi [\mnras] {10.1093/mnras/stx1085}, \href
  {https://ui.adsabs.harvard.edu/#abs/2017MNRAS.469.3688D} {469, 3688}

\bibitem[\protect\citeauthoryear{{Dziembowski}}{{Dziembowski}}{2016}]{2016CoKon.105...23D}
{Dziembowski} W.~A.,  2016, Commmunications of the Konkoly Observatory Hungary,
  \href {https://ui.adsabs.harvard.edu/#abs/2016CoKon.105...23D} {105, 23}

\bibitem[\protect\citeauthoryear{{Foster}}{{Foster}}{1995}]{1995AJ....109.1889F}
{Foster} G.,  1995, \mn@doi [\aj] {10.1086/117416}, \href
  {http://adsabs.harvard.edu/abs/1995AJ....109.1889F} {109, 1889}

\bibitem[\protect\citeauthoryear{{Greer}, {Payne}, {Norton}, {Maxted},
  {Smalley}, {West}, {Wheatley}  \& {Kolb}}{{Greer}
  et~al.}{2017}]{2017A&A...607A..11G}
{Greer} P.~A.,  {Payne} S.~G.,  {Norton} A.~J.,  {Maxted} P.~F.~L.,  {Smalley}
  B.,  {West} R.~G.,  {Wheatley} P.~J.,   {Kolb} U.~C.,  2017, \mn@doi [\aap]
  {10.1051/0004-6361/201630296}, \href
  {https://ui.adsabs.harvard.edu/#abs/2017A&A...607A..11G} {607, A11}

\bibitem[\protect\citeauthoryear{{Gruberbauer} et~al.,}{{Gruberbauer}
  et~al.}{2007}]{2007MNRAS.379.1498G}
{Gruberbauer} M.,  et~al., 2007, \mn@doi [\mnras]
  {10.1111/j.1365-2966.2007.12042.x}, \href
  {https://ui.adsabs.harvard.edu/#abs/2007MNRAS.379.1498G} {379, 1498}

\bibitem[\protect\citeauthoryear{{Hippke}}{{Hippke}}{2015}]{2015ApJ...806...51H}
{Hippke} M.,  2015, \mn@doi [\apj] {10.1088/0004-637X/806/1/51}, \href
  {https://ui.adsabs.harvard.edu/#abs/2015ApJ...806...51H} {806, 51}

\bibitem[\protect\citeauthoryear{{Hippke} \& {Angerhausen}}{{Hippke} \&
  {Angerhausen}}{2015}]{2015ApJ...811....1H}
{Hippke} M.,  {Angerhausen} D.,  2015, \mn@doi [\apj]
  {10.1088/0004-637X/811/1/1}, \href
  {https://ui.adsabs.harvard.edu/#abs/2015ApJ...811....1H} {811, 1}

\bibitem[\protect\citeauthoryear{{Hippke}, {Learned}, {Zee}, {Edmondson},
  {Lindner}, {Kia}, {Ditto}  \& {Stevens}}{{Hippke}
  et~al.}{2015}]{2015ApJ...798...42H}
{Hippke} M.,  {Learned} J.~G.,  {Zee} A.,  {Edmondson} W.~H.,  {Lindner} J.~F.,
   {Kia} B.,  {Ditto} W.~L.,   {Stevens} I.~R.,  2015, \mn@doi [\apj]
  {10.1088/0004-637X/798/1/42}, \href
  {https://ui.adsabs.harvard.edu/#abs/2015ApJ...798...42H} {798, 42}

\bibitem[\protect\citeauthoryear{{Holl} et~al.,}{{Holl}
  et~al.}{2018}]{2018arXiv180409373H}
{Holl} B.,  et~al., 2018, preprint, \href
  {https://ui.adsabs.harvard.edu/#abs/2018arXiv180409373H} {p.
  arXiv:1804.09373} (\mn@eprint {arXiv} {1804.09373})

\bibitem[\protect\citeauthoryear{{Juh{\'a}sz} \& {Moln{\'a}r}}{{Juh{\'a}sz} \&
  {Moln{\'a}r}}{2018}]{2018arXiv180310028J}
{Juh{\'a}sz} {\'A}.~L.,  {Moln{\'a}r} L.,  2018, preprint, \href
  {https://ui.adsabs.harvard.edu/#abs/2018arXiv180310028J} {p.
  arXiv:1803.10028} (\mn@eprint {arXiv} {1803.10028})

\bibitem[\protect\citeauthoryear{{Jurcsik} et~al.,}{{Jurcsik}
  et~al.}{2009}]{2009MNRAS.400.1006J}
{Jurcsik} J.,  et~al., 2009, \mn@doi [\mnras]
  {10.1111/j.1365-2966.2009.15515.x}, \href
  {https://ui.adsabs.harvard.edu/#abs/2009MNRAS.400.1006J} {400, 1006}

\bibitem[\protect\citeauthoryear{{Jurcsik} et~al.,}{{Jurcsik}
  et~al.}{2015}]{2015ApJS..219...25J}
{Jurcsik} J.,  et~al., 2015, \mn@doi [The Astrophysical Journal Supplement
  Series] {10.1088/0067-0049/219/2/25}, \href
  {https://ui.adsabs.harvard.edu/#abs/2015ApJS..219...25J} {219, 25}

\bibitem[\protect\citeauthoryear{Kaplan}{Kaplan}{1964}]{kaplan1964conduct}
Kaplan A.,  1964, The conduct of inquiry: methodology for behavioral science.
Chandler publications in anthropology and sociology, Chandler Pub. Co., \url
  {https://books.google.de/books?id=kOg7AAAAIAAJ}

\bibitem[\protect\citeauthoryear{{Kim}, {Protopapas}, {Bailer-Jones}, {Byun},
  {Chang}, {Marquette}  \& {Shin}}{{Kim} et~al.}{2014}]{2014A&A...566A..43K}
{Kim} D.-W.,  {Protopapas} P.,  {Bailer-Jones} C. A.~L.,  {Byun} Y.-I.,
  {Chang} S.-W.,  {Marquette} J.-B.,   {Shin} M.-S.,  2014, \mn@doi [\aap]
  {10.1051/0004-6361/201323252}, \href
  {https://ui.adsabs.harvard.edu/#abs/2014A&A...566A..43K} {566, A43}

\bibitem[\protect\citeauthoryear{{Kovacs}}{{Kovacs}}{2018}]{2018A&A...614L...4K}
{Kovacs} G.,  2018, \mn@doi [\aap] {10.1051/0004-6361/201833181}, \href
  {https://ui.adsabs.harvard.edu/#abs/2018A&A...614L...4K} {614, L4}

\bibitem[\protect\citeauthoryear{{Leavitt} \& {Pickering}}{{Leavitt} \&
  {Pickering}}{1912}]{1912HarCi.173....1L}
{Leavitt} H.~S.,  {Pickering} E.~C.,  1912, Harvard College Observatory
  Circular, \href {https://ui.adsabs.harvard.edu/#abs/1912HarCi.173....1L}
  {173, 1}

\bibitem[\protect\citeauthoryear{{Lenz} \& {Breger}}{{Lenz} \&
  {Breger}}{2004}]{2004IAUS..224..786L}
{Lenz} P.,  {Breger} M.,  2004, in The A-Star Puzzle. pp 786--790,
  \mn@doi{10.1017/S1743921305009750}

\bibitem[\protect\citeauthoryear{{Lindner}, {Kohar}, {Kia}, {Hippke}, {Learned}
   \& {Ditto}}{{Lindner} et~al.}{2015}]{2015PhRvL.114e4101L}
{Lindner} J.~F.,  {Kohar} V.,  {Kia} B.,  {Hippke} M.,  {Learned} J.~G.,
  {Ditto} W.~L.,  2015, \mn@doi [\prl] {10.1103/PhysRevLett.114.054101}, \href
  {https://ui.adsabs.harvard.edu/#abs/2015PhRvL.114e4101L} {114, 054101}

\bibitem[\protect\citeauthoryear{{Lindner}, {Kohar}, {Kia}, {Hippke}, {Learned}
   \& {Ditto}}{{Lindner} et~al.}{2016}]{2016PhyD..316...16L}
{Lindner} J.~F.,  {Kohar} V.,  {Kia} B.,  {Hippke} M.,  {Learned} J.~G.,
  {Ditto} W.~L.,  2016, \mn@doi [Physica D Nonlinear Phenomena]
  {10.1016/j.physd.2015.10.006}, \href
  {https://ui.adsabs.harvard.edu/#abs/2016PhyD..316...16L} {316, 16}

\bibitem[\protect\citeauthoryear{{Mateu}, {Vivas}, {Downes}, {Brice{\~n}o},
  {Zinn}  \& {Cruz-Diaz}}{{Mateu} et~al.}{2012}]{2012MNRAS.427.3374M}
{Mateu} C.,  {Vivas} A.~K.,  {Downes} J.~J.,  {Brice{\~n}o} C.,  {Zinn} R.,
  {Cruz-Diaz} G.,  2012, \mn@doi [\mnras] {10.1111/j.1365-2966.2012.21968.x},
  \href {https://ui.adsabs.harvard.edu/#abs/2012MNRAS.427.3374M} {427, 3374}

\bibitem[\protect\citeauthoryear{{Minniti} et~al.,}{{Minniti}
  et~al.}{2017}]{2017AJ....153..179M}
{Minniti} D.,  et~al., 2017, \mn@doi [\aj] {10.3847/1538-3881/aa5be4}, \href
  {http://adsabs.harvard.edu/abs/2017AJ....153..179M} {153, 179}

\bibitem[\protect\citeauthoryear{{Moln{\'a}r}}{{Moln{\'a}r}}{2018}]{2018arXiv180310026M}
{Moln{\'a}r} L.,  2018, preprint, \href
  {https://ui.adsabs.harvard.edu/#abs/2018arXiv180310026M} {p.
  arXiv:1803.10026} (\mn@eprint {arXiv} {1803.10026})

\bibitem[\protect\citeauthoryear{{Moln{\'a}r} et~al.,}{{Moln{\'a}r}
  et~al.}{2015}]{2015MNRAS.452.4283M}
{Moln{\'a}r} L.,  et~al., 2015, \mn@doi [\mnras] {10.1093/mnras/stv1638}, \href
  {https://ui.adsabs.harvard.edu/#abs/2015MNRAS.452.4283M} {452, 4283}

\bibitem[\protect\citeauthoryear{{Moln{\'a}r} et~al.,}{{Moln{\'a}r}
  et~al.}{2017}]{2017EPJWC.16004008M}
{Moln{\'a}r} L.,  et~al., 2017, in European Physical Journal Web of
  Conferences. p. 04008 (\mn@eprint {arXiv} {1703.02420}),
  \mn@doi{10.1051/epjconf/201716004008}

\bibitem[\protect\citeauthoryear{{Moskalik} \& {Kolaczkowski}}{{Moskalik} \&
  {Kolaczkowski}}{2008}]{2008CoAst.157..343M}
{Moskalik} P.,  {Kolaczkowski} Z.,  2008, Communications in Asteroseismology,
  \href {https://ui.adsabs.harvard.edu/#abs/2008CoAst.157..343M} {157, 343}

\bibitem[\protect\citeauthoryear{{Moskalik} \& {Ko{\l}aczkowski}}{{Moskalik} \&
  {Ko{\l}aczkowski}}{2009}]{2009MNRAS.394.1649M}
{Moskalik} P.,  {Ko{\l}aczkowski} Z.,  2009, \mn@doi [\mnras]
  {10.1111/j.1365-2966.2009.14428.x}, \href
  {https://ui.adsabs.harvard.edu/#abs/2009MNRAS.394.1649M} {394, 1649}

\bibitem[\protect\citeauthoryear{{Moskalik} \& {Poretti}}{{Moskalik} \&
  {Poretti}}{2003}]{2003A&A...398..213M}
{Moskalik} P.,  {Poretti} E.,  2003, \mn@doi [\aap]
  {10.1051/0004-6361:20021595}, \href
  {https://ui.adsabs.harvard.edu/#abs/2003A&A...398..213M} {398, 213}

\bibitem[\protect\citeauthoryear{{Moskalik} et~al.,}{{Moskalik}
  et~al.}{2015}]{2015MNRAS.447.2348M}
{Moskalik} P.,  et~al., 2015, \mn@doi [\mnras] {10.1093/mnras/stu2561}, \href
  {https://ui.adsabs.harvard.edu/#abs/2015MNRAS.447.2348M} {447, 2348}

\bibitem[\protect\citeauthoryear{{Nemec} et~al.,}{{Nemec}
  et~al.}{2011}]{2011MNRAS.417.1022N}
{Nemec} J.~M.,  et~al., 2011, \mn@doi [\mnras]
  {10.1111/j.1365-2966.2011.19317.x}, \href
  {https://ui.adsabs.harvard.edu/#abs/2011MNRAS.417.1022N} {417, 1022}

\bibitem[\protect\citeauthoryear{{Nemec}, {Cohen}, {Ripepi}, {Derekas},
  {Moskalik}, {Sesar}, {Chadid}  \& {Bruntt}}{{Nemec}
  et~al.}{2013}]{2013ApJ...773..181N}
{Nemec} J.~M.,  {Cohen} J.~G.,  {Ripepi} V.,  {Derekas} A.,  {Moskalik} P.,
  {Sesar} B.,  {Chadid} M.,   {Bruntt} H.,  2013, \mn@doi [\apj]
  {10.1088/0004-637X/773/2/181}, \href
  {https://ui.adsabs.harvard.edu/#abs/2013ApJ...773..181N} {773, 181}

\bibitem[\protect\citeauthoryear{{Netzel}, {Smolec}  \& {Moskalik}}{{Netzel}
  et~al.}{2015a}]{2015MNRAS.447.1173N}
{Netzel} H.,  {Smolec} R.,   {Moskalik} P.,  2015a, \mn@doi [\mnras]
  {10.1093/mnras/stu2409}, \href
  {https://ui.adsabs.harvard.edu/#abs/2015MNRAS.447.1173N} {447, 1173}

\bibitem[\protect\citeauthoryear{{Netzel}, {Smolec}  \& {Dziembowski}}{{Netzel}
  et~al.}{2015b}]{2015MNRAS.451L..25N}
{Netzel} H.,  {Smolec} R.,   {Dziembowski} W.,  2015b, \mn@doi [\mnras]
  {10.1093/mnrasl/slv062}, \href
  {https://ui.adsabs.harvard.edu/#abs/2015MNRAS.451L..25N} {451, L25}

\bibitem[\protect\citeauthoryear{{Netzel}, {Smolec}  \& {Moskalik}}{{Netzel}
  et~al.}{2015c}]{2015MNRAS.453.2022N}
{Netzel} H.,  {Smolec} R.,   {Moskalik} P.,  2015c, \mn@doi [\mnras]
  {10.1093/mnras/stv1758}, \href
  {https://ui.adsabs.harvard.edu/#abs/2015MNRAS.453.2022N} {453, 2022}

\bibitem[\protect\citeauthoryear{{Netzel}, {Smolec}, {Soszy{\'n}ski}  \&
  {Udalski}}{{Netzel} et~al.}{2018}]{2018MNRAS.tmp.1796N}
{Netzel} H.,  {Smolec} R.,  {Soszy{\'n}ski} I.,   {Udalski} A.,  2018, \mn@doi
  [\mnras] {10.1093/mnras/sty1883}, \href
  {https://ui.adsabs.harvard.edu/#abs/2018MNRAS.tmp.1796N} {p.~1796}

\bibitem[\protect\citeauthoryear{{Olech} \& {Moskalik}}{{Olech} \&
  {Moskalik}}{2009}]{2009A&A...494L..17O}
{Olech} A.,  {Moskalik} P.,  2009, \mn@doi [\aap]
  {10.1051/0004-6361:200811441}, \href
  {https://ui.adsabs.harvard.edu/#abs/2009A&A...494L..17O} {494, L17}

\bibitem[\protect\citeauthoryear{{Petersen}}{{Petersen}}{1973}]{1973A&A....27...89P}
{Petersen} J.~O.,  1973, \aap, \href
  {https://ui.adsabs.harvard.edu/#abs/1973A&A....27...89P} {27, 89}

\bibitem[\protect\citeauthoryear{{Plachy}, {Klagyivik}, {Moln{\'a}r},
  {S{\'o}dor}  \& {Szab{\'o}}}{{Plachy} et~al.}{2017}]{2017EPJWC.16004009P}
{Plachy} E.,  {Klagyivik} P.,  {Moln{\'a}r} L.,  {S{\'o}dor} {\'A}.,
  {Szab{\'o}} R.,  2017, in European Physical Journal Web of Conferences. p.
  04009, \mn@doi{10.1051/epjconf/201716004009}

\bibitem[\protect\citeauthoryear{{Plachy}, {Moln{\'a}r}, {B{\'o}di}, {Skarka},
  {Juh{\'a}sz}, {S{\'o}dor}, {Klagyivik}  \& {Szab{\'o}}}{{Plachy}
  et~al.}{2018}]{2018arXiv180705369P}
{Plachy} E.,  {Moln{\'a}r} L.,  {B{\'o}di} A.,  {Skarka} M.,  {Juh{\'a}sz}
  {\'A}.~L.,  {S{\'o}dor} {\'A}.,  {Klagyivik} P.,   {Szab{\'o}} R.,  2018,
  preprint, \href {https://ui.adsabs.harvard.edu/#abs/2018arXiv180705369P} {p.
  arXiv:1807.05369} (\mn@eprint {arXiv} {1807.05369})

\bibitem[\protect\citeauthoryear{{Prudil}, {Smolec}, {Skarka}  \&
  {Netzel}}{{Prudil} et~al.}{2017}]{2017MNRAS.465.4074P}
{Prudil} Z.,  {Smolec} R.,  {Skarka} M.,   {Netzel} H.,  2017, \mn@doi [\mnras]
  {10.1093/mnras/stw3010}, \href
  {https://ui.adsabs.harvard.edu/#abs/2017MNRAS.465.4074P} {465, 4074}

\bibitem[\protect\citeauthoryear{{Rauer} et~al.,}{{Rauer}
  et~al.}{2014}]{2014ExA....38..249R}
{Rauer} H.,  et~al., 2014, \mn@doi [Experimental Astronomy]
  {10.1007/s10686-014-9383-4}, \href
  {https://ui.adsabs.harvard.edu/#abs/2014ExA....38..249R} {38, 249}

\bibitem[\protect\citeauthoryear{{Reegen}}{{Reegen}}{2007}]{2007A&A...467.1353R}
{Reegen} P.,  2007, \mn@doi [\aap] {10.1051/0004-6361:20066597}, \href
  {http://adsabs.harvard.edu/abs/2007A%26A...467.1353R} {467, 1353}

\bibitem[\protect\citeauthoryear{{Ricker} et~al.,}{{Ricker}
  et~al.}{2015}]{2015JATIS...1a4003R}
{Ricker} G.~R.,  et~al., 2015, \mn@doi [Journal of Astronomical Telescopes,
  Instruments, and Systems] {10.1117/1.JATIS.1.1.014003}, \href
  {https://ui.adsabs.harvard.edu/#abs/2015JATIS...1a4003R} {1, 014003}

\bibitem[\protect\citeauthoryear{{Sesar} et~al.,}{{Sesar}
  et~al.}{2013}]{2013AJ....146...21S}
{Sesar} B.,  et~al., 2013, \mn@doi [\aj] {10.1088/0004-6256/146/2/21}, \href
  {http://adsabs.harvard.edu/abs/2013AJ....146...21S} {146, 21}

\bibitem[\protect\citeauthoryear{{Sheets} \& {Deming}}{{Sheets} \&
  {Deming}}{2014}]{2014ApJ...794..133S}
{Sheets} H.~A.,  {Deming} D.,  2014, \mn@doi [\apj]
  {10.1088/0004-637X/794/2/133}, \href
  {https://ui.adsabs.harvard.edu/#abs/2014ApJ...794..133S} {794, 133}

\bibitem[\protect\citeauthoryear{{Sheets} \& {Deming}}{{Sheets} \&
  {Deming}}{2017}]{2017AJ....154..160S}
{Sheets} H.~A.,  {Deming} D.,  2017, \mn@doi [\aj] {10.3847/1538-3881/aa88b9},
  \href {https://ui.adsabs.harvard.edu/#abs/2017AJ....154..160S} {154, 160}

\bibitem[\protect\citeauthoryear{{Smolec} \& {{\'S}niegowska}}{{Smolec} \&
  {{\'S}niegowska}}{2016}]{2016MNRAS.458.3561S}
{Smolec} R.,  {{\'S}niegowska} M.,  2016, \mn@doi [\mnras]
  {10.1093/mnras/stw553}, \href
  {https://ui.adsabs.harvard.edu/#abs/2016MNRAS.458.3561S} {458, 3561}

\bibitem[\protect\citeauthoryear{{Smolec}, {Prudil}, {Skarka}  \&
  {Bakowska}}{{Smolec} et~al.}{2016}]{2016MNRAS.461.2934S}
{Smolec} R.,  {Prudil} Z.,  {Skarka} M.,   {Bakowska} K.,  2016, \mn@doi
  [\mnras] {10.1093/mnras/stw1519}, \href
  {https://ui.adsabs.harvard.edu/#abs/2016MNRAS.461.2934S} {461, 2934}

\bibitem[\protect\citeauthoryear{{Smolec}, {Dziembowski}, {Moskalik}, {Netzel},
  {Prudil}, {Skarka}  \& {Soszynski}}{{Smolec}
  et~al.}{2017a}]{2017EPJWC.15206003S}
{Smolec} R.,  {Dziembowski} W.,  {Moskalik} P.,  {Netzel} H.,  {Prudil} Z.,
  {Skarka} M.,   {Soszynski} I.,  2017a, in European Physical Journal Web of
  Conferences. p. 06003, \mn@doi{10.1051/epjconf/201715206003}

\bibitem[\protect\citeauthoryear{{Smolec}, {Moskalik}, {Ka{\l}u{\.z}ny},
  {Pych}, {R{\'o}{\.z}yczka}  \& {Thompson}}{{Smolec}
  et~al.}{2017b}]{2017MNRAS.467.2349S}
{Smolec} R.,  {Moskalik} P.,  {Ka{\l}u{\.z}ny} J.,  {Pych} W.,
  {R{\'o}{\.z}yczka} M.,   {Thompson} I.~B.,  2017b, \mn@doi [\mnras]
  {10.1093/mnras/stx088}, \href
  {https://ui.adsabs.harvard.edu/#abs/2017MNRAS.467.2349S} {467, 2349}

\bibitem[\protect\citeauthoryear{{Soszy{\'n}ski} et~al.,}{{Soszy{\'n}ski}
  et~al.}{2009}]{2009AcA....59....1S}
{Soszy{\'n}ski} I.,  et~al., 2009, \actaa, \href
  {https://ui.adsabs.harvard.edu/#abs/2009AcA....59....1S} {59, 1}

\bibitem[\protect\citeauthoryear{{S{\"u}veges} \& {Anderson}}{{S{\"u}veges} \&
  {Anderson}}{2018a}]{2018MNRAS.478.1425S}
{S{\"u}veges} M.,  {Anderson} R.~I.,  2018a, \mn@doi [\mnras]
  {10.1093/mnras/sty891}, \href
  {https://ui.adsabs.harvard.edu/#abs/2018MNRAS.478.1425S} {478, 1425}

\bibitem[\protect\citeauthoryear{{S{\"u}veges} \& {Anderson}}{{S{\"u}veges} \&
  {Anderson}}{2018b}]{2018A&A...610A..86S}
{S{\"u}veges} M.,  {Anderson} R.~I.,  2018b, \mn@doi [\aap]
  {10.1051/0004-6361/201628870}, \href
  {https://ui.adsabs.harvard.edu/#abs/2018A&A...610A..86S} {610, A86}

\bibitem[\protect\citeauthoryear{{S{\"u}veges} et~al.,}{{S{\"u}veges}
  et~al.}{2012}]{2012MNRAS.424.2528S}
{S{\"u}veges} M.,  et~al., 2012, \mn@doi [\mnras]
  {10.1111/j.1365-2966.2012.21229.x}, \href
  {https://ui.adsabs.harvard.edu/#abs/2012MNRAS.424.2528S} {424, 2528}

\bibitem[\protect\citeauthoryear{{Szab{\'o}} et~al.,}{{Szab{\'o}}
  et~al.}{2014}]{2014A&A...570A.100S}
{Szab{\'o}} R.,  et~al., 2014, \mn@doi [\aap] {10.1051/0004-6361/201424522},
  \href {https://ui.adsabs.harvard.edu/#abs/2014A&A...570A.100S} {570, A100}

\bibitem[\protect\citeauthoryear{{Teachey}, {Kipping}  \& {Schmitt}}{{Teachey}
  et~al.}{2018}]{2018AJ....155...36T}
{Teachey} A.,  {Kipping} D.~M.,   {Schmitt} A.~R.,  2018, \mn@doi [\aj]
  {10.3847/1538-3881/aa93f2}, \href
  {https://ui.adsabs.harvard.edu/#abs/2018AJ....155...36T} {155, 36}

\bibitem[\protect\citeauthoryear{{Tyson}}{{Tyson}}{2002}]{2002SPIE.4836...10T}
{Tyson} J.~A.,  2002, in Survey and Other Telescope Technologies and
  Discoveries. pp 10--20 (\mn@eprint {arXiv} {astro-ph/0302102}),
  \mn@doi{10.1117/12.456772}

\bibitem[\protect\citeauthoryear{{Udalski}, {Szymanski}, {Soszynski}  \&
  {Poleski}}{{Udalski} et~al.}{2008}]{2008AcA....58...69U}
{Udalski} A.,  {Szymanski} M.~K.,  {Soszynski} I.,   {Poleski} R.,  2008,
  \actaa, \href {https://ui.adsabs.harvard.edu/#abs/2008AcA....58...69U} {58,
  69}

\bibitem[\protect\citeauthoryear{{Udalski}, {Szyma{\'n}ski}  \&
  {Szyma{\'n}ski}}{{Udalski} et~al.}{2015}]{2015AcA....65....1U}
{Udalski} A.,  {Szyma{\'n}ski} M.~K.,   {Szyma{\'n}ski} G.,  2015, \actaa,
  \href {https://ui.adsabs.harvard.edu/#abs/2015AcA....65....1U} {65, 1}

\bibitem[\protect\citeauthoryear{{Vivas} \& {Zinn}}{{Vivas} \&
  {Zinn}}{2006}]{2006AJ....132..714V}
{Vivas} A.~K.,  {Zinn} R.,  2006, \mn@doi [\aj] {10.1086/505200}, \href
  {https://ui.adsabs.harvard.edu/#abs/2006AJ....132..714V} {132, 714}

\bibitem[\protect\citeauthoryear{{Vivas} et~al.,}{{Vivas}
  et~al.}{2004}]{2004AJ....127.1158V}
{Vivas} A.~K.,  et~al., 2004, \mn@doi [\aj] {10.1086/380929}, \href
  {https://ui.adsabs.harvard.edu/#abs/2004AJ....127.1158V} {127, 1158}

\bibitem[\protect\citeauthoryear{{Zinn}, {Horowitz}, {Vivas}, {Baltay},
  {Ellman}, {Hadjiyska}, {Rabinowitz}  \& {Miller}}{{Zinn}
  et~al.}{2014}]{2014ApJ...781...22Z}
{Zinn} R.,  {Horowitz} B.,  {Vivas} A.~K.,  {Baltay} C.,  {Ellman} N.,
  {Hadjiyska} E.,  {Rabinowitz} D.,   {Miller} L.,  2014, \mn@doi [\apj]
  {10.1088/0004-637X/781/1/22}, \href
  {https://ui.adsabs.harvard.edu/#abs/2014ApJ...781...22Z} {781, 22}

\makeatother
\end{thebibliography}
\bsp
\label{lastpage}
\end{document}